\documentclass[preprint,aps,showpacs,preprintnumbers,amsmath,amssymb]{revtex4}

\usepackage[american]{babel}
\usepackage{graphicx}
\usepackage{dcolumn}
\usepackage{bm}

\begin{document}

\preprint{APS/123-QED}

\title{{\it Ab-initio} self-energy corrections in systems
with metallic screening}

\author{Marco Cazzaniga}
 \email{marco.cazzaniga@unimi.it}
\author{Nicola Manini}
\author{Luca Guido Molinari}
\author{Giovanni Onida}
 \affiliation{
Physics Department, Universit\`a degli Studi di Milano, I-20133 Milan (Italy)\\
European Theoretical Spectroscopy Facility (ETSF)
}

\date{November 20, 2007}

\begin{abstract}
The calculation of self-energy corrections to the electron bands of a metal
requires the evaluation of the intraband contribution to the
polarizability in the small-${\bf q}$  limit.
When neglected, as in standard $GW$ codes
for semiconductors and insulators, a spurious gap opens at the Fermi
energy.
Systematic methods to include intraband contributions to the polarizability
exist, but require a computationally intensive Fermi-surface integration.
We propose a numerically cheap and stable method, based 
on a fit of the power expansion of the polarizability in the 
small-${\bf q}$ region.
We test it on the homogeneous electron gas
and on real metals such as sodium and aluminum.
\end{abstract}

\pacs{71.15.-m, 
      71.38.Cn, 
      71.20.-b, 
      71.15.Dx, 
      }

\maketitle

\section{\label{sec:intro}Introduction}
More than 20 years of successful applications have established Hedin's $GW$
approach \cite{Hedin, HedinL} and its numerical implementations 
\cite{gss,Hl}
as the
state-of-the-art and most widely used theoretical method for {\it
ab-initio} bandstructure calculations including self-energy effects.
Efficient algorithms have been devised to encompass the major numerical
bottlenecks in such calculations, e.g. by avoiding {\bf k}-space
convolutions by a space-time method \cite{godby-st}, or avoiding summations
over empty states in the evaluation of the polarizability
\cite{reining-empty, reining-empty2}, or using localized basis functions 
\cite{loc}, and/or  model screening functions \cite{reining-mod,mass,shir2}.
Several computer codes have been devised for ab-initio $GW$ calculations,
and are presently available,
often under public domain \cite{ABINIT,SaX,self}.
However, systems with metallic screening present an additional, numerically
challenging, difficulty: in the evaluation of the {\bf k}-space integrals
for the intraband contribution to the electron screening, the contribution
of the Fermi surface can dramatically slow down the convergence with
respect to the {\bf k}-space sampling.
As a consequence, the possibility to perform such {\it ab-initio} $GW$
calculations in gapless systems with a large unit cell is hindered. Even
worse, when calculations are performed with standard computer codes,
unconverged {\bf k}-point sampling gives rise to a spurious gap at
the Fermi level.
The gap vanishes only in the limit of infinitely dense
sampling, and is shown to close very slowly as the number of {\bf k}-points
increases.
Solutions based on explicit Fermi surface integration \cite{Mak, LeeChang,
Marini2} are effective but result in cumbersome coding and substantial
increase of computation time.

In this paper we present a numerically stable and efficient method, based
on a Taylor expansion of the polarizability matrix in the small-{\bf q} 
region, which includes intraband contributions and avoids explicit 
Fermi-surface calculations.
The method has been implemented successfully into the \verb;abinit;
\cite{ABINIT,Gonze05} package, and is shown to remove the spurious gap at
the Fermi level already with a limited number of {\bf k}-points.
Results are presented for the homogeneous electron gas (HEG) as well as for 
real metals such as Na and Al.

This paper is organized as follows: in Sec.\ \ref{sec:theory} we briefly
review the standard $GW$ scheme and describe the difficulties that arise when
it is applied to metals naively.
In Sec.\ \ref{sec:met} we analyze the origin of the problem, and propose  
our solution in Sec.\ \ref{sec:fit}.
In Sec.\ \ref{sec:res} we test the method on different metallic
systems,  and we discuss the results in Sec.\ \ref{sec:conc}.

\section{\label{sec:theory}Theoretical framework}

The present work deals with the many-body problem in the standard
Hedin's scheme based on the following set of self-consistent 
equations \cite{Hedin}:
\begin{eqnarray}
\label{eq:1hedin}
G(1,2)&=&G_0(1,2)+\!\!\!{\int}G_0(1,3)\Sigma(3,4)G(4,2)\,d3d4\\
\label{eq:2hedin}
\Gamma(1,2;3)&=&\delta(1,2)\delta(1,3)
+\int
\frac{\delta\Sigma(1,2)}{{\delta}G(4,5)}G(4,6)G(7,5)\Gamma(6,7;3)\,d4d5d6d7\\ 
\label{eq:3hedin}
\chi(1,2)&=&-i{\int}G(1,3)G(4,1)\Gamma(3,4;2)\,d3d4\\
\label{eq:4hedin}
W(1,2)&=&v_C(1,2)+{\int}v_C(1,3)\chi(3,4)W(4,2)\,d3d4\\
\label{eq:5hedin}
\Sigma(1,2)&=&i{\int}G(1,3)W(4,1)\Gamma(3,2;4)\,d3d4
\,,
\end{eqnarray} 
$G$ and $G_0$ are the exact and Hartree's Green functions for the electron, 
$v_C$ is the bare Coulomb interaction, $W$ is the screened potential, 
$\chi$ is the electric polarizability, $\Sigma$ is the self energy, 
and $\Gamma$ is the vertex function.
An argument such as "$1$" stands for the set of position, time and spin
variables $({\bf r}_1,t_1,\sigma_1)$.
Equations (\ref{eq:1hedin}-\ref{eq:5hedin}) constitute a formally closed
set of equations for the five correlators.
The functional derivative in Eq.\ (\ref{eq:2hedin}) provides the vertex 
corrections and is a formidable computational difficulty. The most important 
approximation that is usually made is to neglect the vertex entirely 
and put $\Gamma(1,2;3)=\delta(1,2)\delta(1,3)$ in the remaining four
equations. This explains the name $GW$, since Eq.~(\ref{eq:5hedin}) now 
simplifies to the product 
\begin{equation}\label{SigmaGW}
\Sigma(1,2)=iG(1,2)W(2,1)\,.
\end{equation}

A $GW$ calculation
proceeds as follows. One assumes initially $\Sigma = 0$, $G=G_0$ in 
Eq.~(\ref{eq:1hedin}). Next, one determines $\chi = \chi_0$ through 
Eq.~(\ref{eq:3hedin}) with $\Gamma = 1$,
and computes $W$ from Eq.~(\ref{eq:4hedin}). 
The first estimate of $\Sigma$ is obtained in Eq.~(\ref{SigmaGW}), and can be 
used to update $G$ and the other correlators. This procedure can be iterated 
until self-consistency is reached.
However, several non-selfconsistent $GW$ approaches are possible \cite{Fabien}.
In the present work, we choose to perform calculations within the so-called
$G_0W_0$ approximation \cite{gss,Hl}, which stops the iteration without
updating $G$ and $W$.
One first evaluates the independent-particle polarizability
\begin{equation}\label{eq:chi0}
\chi_0(1,2)=-i \, G_0(1,2)\,G_0(2,1)
\end{equation}
and the dielectric function
\begin{equation}\label{eq:W0}
\epsilon(1,2)=\delta (1,2)-\int v_C(1,3)\,\chi_0(3,2)\,d3d4
\,,
\end{equation}
which provides the solution of Eq.~\eqref{eq:4hedin}, $W_0(1,2)=\int
\epsilon^{-1}(1,3)\,v_C(3,2)\,d3$, and the self energy
$\Sigma(1,2)=iG_0(1,2)W_0(2,1)$.
This approximation is usually a successful one, while self-consistent $GW$
has been shown to lead to a worse treatment of electron correlations in
prototypical systems such as the HEG (where it gives a bandwidth larger
than the DFT one \cite{vonBarth}) and solid silicon (where the band gap
turns out even larger than experiment \cite{scfsolid}).
The simpler $G_0W_0$ approach leads typically to a 10\% bandwidth reduction
with respect to DFT, in better agreement with experiment, thus suggesting a
partial cancellation of errors due to lack of self-consistency and of
vertex corrections in \cite{MS,shir}.

In the $G_0W_0$ approximation, one can start from a DFT-LDA
electronic-structure calculation. The quasiparticle energies $E_j$ are
hence evaluated as first order corrections to the Kohn-Sham (KS) eigenvalues
$\epsilon_j$, with respect to the perturbation $(\Sigma -V_{XC})$, and by
linearizing the energy dependence of $\Sigma$ \cite{gss,Hl}:
\begin{eqnarray}\label{eig:eq}
E_j\simeq\epsilon_j+\frac{\langle\Sigma(\epsilon_j) -
  V_{XC}\rangle}{1-\left\langle
  \frac{\partial\Sigma(\omega)}{\partial\omega}
  \big|_{\omega=\epsilon_j}\right\rangle} 
\,.
\end{eqnarray} 
Expectation values are taken on the corresponding KS state $|{\bf
k},j\rangle$; the denominator is the quasiparticle weight.
One of the heaviest parts of the $G_0W_0$ computation is the inversion of
the symmetrized dielectric matrix, which in reciprocal space reads:
\begin{eqnarray}\label{eq:epsrecip}
\epsilon_{{\bf G},{\bf G'}}({\bf q}, \omega) = \delta_{{\bf G},{\bf G'}}-4\pi
\frac{1}{|{\bf q}+{\bf G}|}\,\chi_{0\,{\bf G,G'}}({\bf q},\omega)\,
\frac{1}{|{\bf q}+{\bf G'}|}
\,.
\end{eqnarray}
The inversion must be performed on a mesh of frequencies spanning a range 
significantly wider than the range of interest for the bandstructure. 

The inverse dielectric matrix leads immediately to the effective screened
potential:
\begin{eqnarray}
W_{{\bf G},{\bf G'}}({\bf q},\omega) = 4\pi\frac{1}{|{\bf q}+{\bf G}|}
\epsilon^{-1}_{{\bf G},{\bf G'}}({\bf q},\omega)\frac{1}{|{\bf q}+{\bf G'}|}
\,.
\end{eqnarray}
A great simplification can be achieved by introducing an additional
plasmon-pole approximation, where the frequency dependence of each ${\bf
G}$,${\bf G}^\prime$ matrix element is parameterized by:
\begin{eqnarray}
\epsilon^{-1}_{{\bf G},{\bf G}'}({\bf q},\omega)=\delta_{{\bf G},{\bf G}'}+
\frac {\Omega^2_{{\bf G},{\bf G}'}({\bf q})}{\omega^2 -
	\tilde\omega^2_{{\bf G},{\bf G}'}({\bf q})}
\,.
\end{eqnarray}
The parameters $\Omega^2_{{\bf G},{\bf G'}}({\bf q})$ and $\tilde
\omega^2_{{\bf G},{\bf G'}}({\bf q})$ are determined by evaluating the
polarizability $\chi_{0\,{\bf G},{\bf G'}}({\bf q},\omega)$ only at two
values of the frequency, usually at $\omega=0$ and at a purely imaginary
frequency of the magnitude of the plasma frequency $\omega=i\omega_P$.
In the following we adopt this plasmon-pole model, since the difficulties
related to the small wave-vector screening
would occur identically if the detailed $\omega$ dependence 
of $\epsilon^{-1}$ were considered.

The polarizability $\chi_0$ is given by the standard expression
\cite{Adler62,Wiser63}
\begin{eqnarray}\label{eq:chi}
\chi_{0\,{\bf G},{\bf G}'}({\bf q},i\omega)
&=&
-\frac{2}{V_{BZ}}\sum_{j,j'}\int_{BZ} d^3k\,
\frac{f(\epsilon_{j'}({\bf k}+{\bf q})) - f(\epsilon_j({\bf k})) }
{i\omega-[\epsilon_{j'}({\bf k}+{\bf q})-\epsilon_j({\bf k})]}
\nonumber \\
&&\langle{\bf k},j|e^{-i({\bf q}+{\bf G})\cdot{\bf \hat r}}
|{\bf k}+{\bf q},j'\rangle
\langle{\bf k}+{\bf q},j'|e^{i({\bf q}+{\bf G}')\cdot{\bf \hat r}}
|{\bf k},j\rangle
\,,
\end{eqnarray}
where $f(\epsilon )$ are Fermi occupation numbers at a small smearing
temperature, $|{\bf k},j\rangle$ are the KS states, and the factor
$2$ accounts for spin.
Complex conjugation gives $\chi_{0\,{\bf G},{\bf G'}}^*({\bf q},i\omega )=
\chi_{0\,{\bf G',G}}({\bf q},i\omega )$, hence also $\epsilon_{{\bf G},{\bf
G'}}$ is a Hermitian matrix for purely imaginary frequencies.
%
For $\omega\neq 0$, ${\bf q}={\bf 0}$ and ${\bf G}$ or ${\bf G}'$ equal to
${\bf 0}$, this expression vanishes exactly because of orthogonality
($j\neq j'$ terms) or equality of Fermi numbers ($j=j'$).
The rate at which $\chi_0$ vanishes as ${\bf q}\to {\bf 0}$ is relevant for
contrasting the Coulomb singularity that appears in the dielectric matrix. 
To take care of interband terms ($j\neq j'$), a standard solution is to
expand the matrix elements of Eq. (\ref{eq:chi}) by means of the formula
\cite{Hyb-lu}:
\begin{eqnarray}\label{eq:exp}
\langle{\bf k},j|
e^{-i{\bf q}\cdot {\bf \hat r}}|{\bf k}+{\bf q},j'\rangle
\mathop{\simeq}_{{\bf q}\rightarrow 0}
\frac{\langle{\bf k},j|+i{\bf q}\cdot\nabla_{\bf r}|{\bf k},j'\rangle+
\langle{\bf k},j|[V_{NL},i{\bf q}\cdot {\bf \hat r}]
|{\bf k},j'\rangle}{\epsilon_{j'}({\bf k})-\epsilon_j({\bf k})}
\,,
\end{eqnarray}
where $V_{NL}$ is the non local part of the pseudopotential.
%
By substituting this expansion into Eq.\ (\ref{eq:chi}) one gets a small
${\bf q}$ expansion of the polarizability, which can be used to evaluate
the ${\bf q} \rightarrow {\bf 0}$ limit of $q^{-2}\,\chi_0({\bf q})$
appearing in Eq.\ \eqref{eq:epsrecip} when ${\bf G}={\bf G}^\prime=0$.
Intraband terms $(j=j')$ are put to zero. 
While this method is satisfactory for semiconductors, it gives rise to
substantial difficulties for metals, where {\em intraband} terms are also
important.
This leads to an incorrect evaluation of the $\chi_0$ contributions in the
${\bf q}$-space region closest to the origin, i.e. at one out of $N_{\rm kpt}$
points of the mesh of ${\bf q}$ points.
At first sight, as $\chi_0$ enters the calculation of $\Sigma$ through a
$N_{\rm kpt}$-discretized ${\bf q}$-convolution in reciprocal space, one might
think that this single incorrect value should affect the energy corrections
$\langle\Sigma - V_{XC}\rangle$, with an error of order
$N_{\rm kpt}^{-1}$.
However, the singular behavior of the Coulomb repulsion $v_C$ near ${\bf
q}={\bf 0}$ requires an explicit integration around the singular point,
which makes the final outcome sensitive to the incorrect $\chi_{0\,{\bf
0},{\bf 0}}({\bf 0},i\omega)$ with an error of order $N_{\rm
kpt}^{-\frac13}$.

\section{\label{sec:met}Naive application of a standard $G_0W_0$
code to metallic systems}

\begin{figure}
\includegraphics[width=8cm]{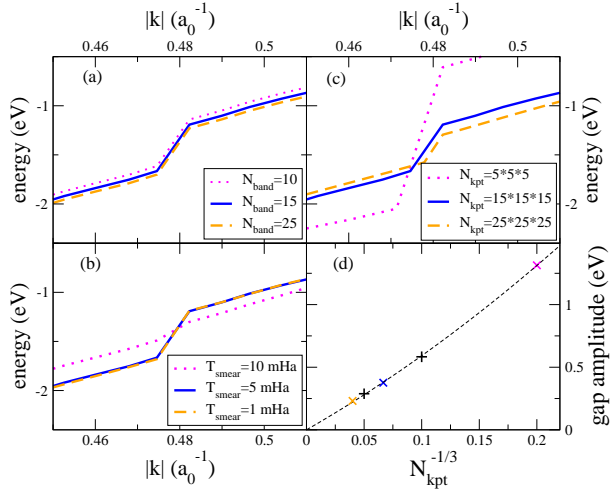}
\caption{\label{fig:conv_na} (Color online)
$G_0W_0$ band structure of Na (110 direction), showing 
the appearance of an unphysical gap, and its dependence on different numerical
convergence parameters. 
Panel (a) shows the dependence with respect to the number of empty states
in Eq.~(\ref{eq:chi}); (b) with respect to the smearing temperature; (c)
with respect to the ${\bf k}$-point mesh.
Panel (d) shows the dependence of the unphysical gap on the inverse number
of ${\bf k}$-points in each direction; the dashed line is a fitted
$a_1N_{\rm kpt}^{-1/3}+a_2N_{\rm kpt}^{-2/3}$.
}
\end{figure}

The incorrect small-${\bf q}$ values of $\chi_0$ induce the opening of an
unphysical gap at the Fermi energy in the $G_0W_0$ band dispersion 
of simple metals (such as the HEG and sodium), as shown in 
Fig.~\ref{fig:conv_na}.
The figure also shows the convergence properties of the width of this
unphysical gap, computed by extrapolation from the two sides.
The only significant dependency is on the number $N_{\rm kpt}$ of sample
points in the ${\bf k}$-space mesh:
Fig.~\ref{fig:conv_na} shows that the unphysical gap does tend to close for
increasing mesh size, but only extremely slowly, as
$N_{\rm kpt}^{-\frac{1}{3}}$, for the reasons discussed at the end of
Sec.~\ref{sec:theory}.
Therefore, it is practically impossible to close the gap by brute-force
mesh refinement, especially because the computation time of the dielectric
matrix grows as $N_{\rm kpt}^2$.

The spurious gap is essentially independent of most numerical convergence
parameters, such as the number of empty states and the smearing
temperature, as shown in Fig.~\ref{fig:conv_na}(a,b).
A larger smearing temperature for electronic occupancy would reduce this
unphysical gap, but it is a mere technical device, and convergence should
be checked in the limit of vanishingly small smearing, where the actual
metallic state is recovered.

\begin{figure}
\includegraphics[width=8cm]{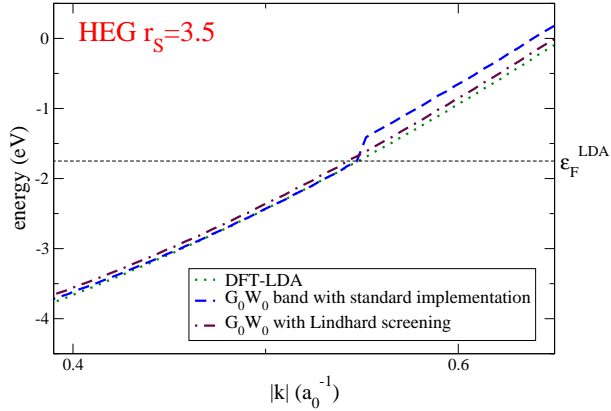}
\caption{\label{fig:bande1} (Color online)
Bandstructure for the HEG ($r_s=3.5\,a_0$) computed
with the standard implementation of the $G_0W_0$ method (dashed line).
The spurious gap, caused by the lack of the intraband term in the
screening, is removed when the computed polarizability is replaced by the
Lindhard one (dot-dashed line).
The KS band is also displayed for reference (dotted line).
}
\end{figure}

\begin{figure}
\includegraphics[width=8cm]{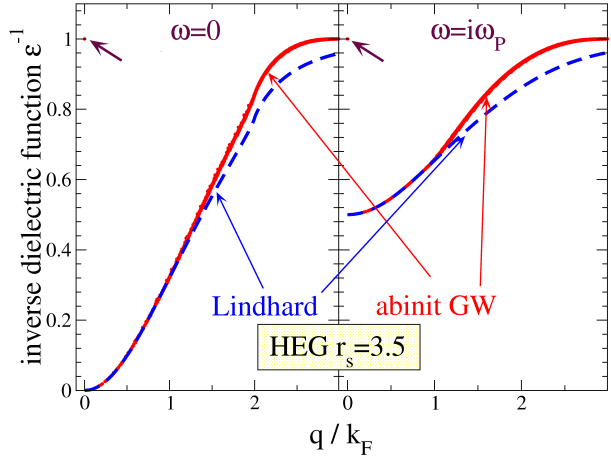}
\caption{\label{fig:jel_eps} (Color online)
Numerically computed HEG  screening function $\epsilon^{-1}({\bf
q},\omega)$ ($r_s=3.5\,a_0$), for $\omega = 0$ and  $\omega = i\omega_P$, 
compared to the Lindhard function.
For ${\bf q}\rightarrow 0$ the discontinuously incorrect points --pointed at
by arrows-- appear, due to the lack of the intraband term. The differences
at large ${\bf q}$ are due to the finite number of empty states included in
the sums of Eq.~(\ref{eq:chi}).
}
\end{figure}

The origin of the unphysical gap is the incorrect ${\bf q}={\bf 0}$ value
of the screening function as demonstrated in Fig.\ \ref{fig:bande1}, where
the gap is shown to disappear when the numerical dielectric matrix is
replaced by the Lindhard function \cite{Mahan}.
In metals, the dielectric function $\epsilon$ is expected to diverge when
both $\omega \rightarrow 0$ and ${\bf q} \rightarrow {\bf 0}$ (by contrast,
it goes to its finite static limit in semiconductors and insulators).
For example for the HEG, interband transitions do not contribute to the sum
in Eq.\ (\ref{eq:chi}).  At the same time, expression (\ref{eq:exp}) cannot
yield correct intraband ($j=j'$) contributions for ${\bf q}={\bf 0}$, and
in practice standard codes do not evaluate such terms due to the equality
of the occupancy factors.
The resulting incorrect null value of $\chi_0({\bf q}\to {\bf 0},i\omega)$
yields $\epsilon^{-1}({\bf q}\to {\bf 0},i\omega)=1$, rather than the
correct $\epsilon^{-1}({\bf q}\to {\bf 0},i\omega)
=\frac{\omega^2}{\omega^2+\omega_P^2}$, as shown in Fig. \ref{fig:jel_eps}
where numerical results are compared with the Lindhard function.

A similar discontinuity in $G_0W_0$ corrections occurs for real metals such
as Na and Al.
Differently from the HEG, we find $\epsilon^{-1}({\bf 0},i\omega)<1$, due
to the nonzero interband contributions \cite{heg}.
In particular we obtain
$\epsilon^{-1}_{\rm Na}({\bf q}={\bf 0},0)\simeq 0.94$, similar to the
incorrect HEG value, and $\epsilon^{-1}_{\rm Al}({\bf q}={\bf
0},0)\simeq 0.008$.
The latter nears the proper Drude value, due to a substantial part of the
aluminum Fermi surface being very close to a Brillouin-zone boundary, thus
putting many of the metallic contributions of Eq.\ (\ref{eq:chi})
effectively into inter- rather than intra-band terms.
For this reason, in the case of Al, the error induced by neglecting the
intraband term is so small that the unphysical gap is almost invisible.

\section{\label{sec:fit}Extrapolated small-$\bf q$ polarizability}

The solution we propose in this paper is devised to avoid the explicit
(numerically expensive) integration over the Fermi surface that would be
required for a straightforward inclusion of the intraband term.
We propose to compute the small ${\bf q}$ polarizability by a fit
of the expected asymptotics.
The time-reversal invariance implies the following symmetry of the matrix
polarization:
\begin{eqnarray}
\chi_{0\,{\bf G},{\bf G}'}({\bf q},\omega)=
\chi_{0\,-{\bf G}',-{\bf G}}(-{\bf q},\omega)
\,.
\end{eqnarray}
Therefore, the small-${\bf q}$ expansion of 
$\chi_{0\,{\bf 0,0}}({\bf q},i\omega)$ includes only even powers.
The expansion of the intraband term $(j=j')$ in Eq.(\ref{eq:chi}) is
\begin{eqnarray}\label{chintra}
\chi^{intra}_{0\,{\bf 0,0}}({\bf q},i\omega)\approx 
\frac2{V_{BZ}}\sum_j\int \! d^3k \, \delta(\mu-\epsilon_j({\bf k}))\,
\frac{{\bf q}\cdot \nabla_{\bf k}\epsilon_j}
{i\omega-{\bf q}\cdot {\nabla}_{\bf k}\epsilon_j} 
|1+{\bf q}\cdot \langle {\bf k},j|\nabla_{\bf k}-i{\bf r}|{\bf k},j\rangle
|^2 
.
\end{eqnarray}
The diagonal matrix element in Eq.~\eqref{chintra} is purely imaginary,
therefore the last factor is 1 plus a ${\bf q}$-quadratic contribution.
For $\omega =0$ the intraband term is then a constant proportional to the
density of states at the Fermi energy, plus quadratic corrections.
For $\omega\neq 0$ the term linear in ${\bf q}$ cancels because
$\nabla_{\bf k}\epsilon_j$ is odd and the integral vanishes: the expansion
begins with quadratic terms.
The expansion of the interband $j\neq j'$ term is easily seen to be no less
than quadratic.
To sum up, we use the following expression:
\begin{equation}
\label{eq:fit0}
\chi_{0\,{\bf 0},{\bf 0}}^{\rm fit}({\bf q},\omega)=
A^\omega+\sum_{rs}B_{rs}^\omega q_r q_s
\,,
\end{equation}
where $A^\omega$, $B_{rs}^\omega$ are real adjustable parameters, and
$A^{i\omega_P}=0$ for $\omega=i\omega_P$.
The matrices $B^\omega$ are symmetric, and may have further symmetries
depending on the crystal geometry.

The off-diagonal elements ${\bf G}={\bf 0}$ ${\bf G'}\neq{\bf 0}$ of
$\chi_0$ (the so-called ``wings'' of the matrix) are affected by a similar
error, since they also contain the contributions of Eq.\ (\ref{eq:exp}).
We also fit the intraband  contribution to 
\begin{eqnarray}
\label{eq:fit0wings}
\chi_{0\,{\bf 0},{\bf G'}}^{\rm fit}({\bf q},\omega)=
C^{\omega\,{\bf G'}}+\sum_{r}D_r^{\omega\,{\bf G'}} q_r
\,,
\end{eqnarray}
where $C^{\omega\,{\bf G'}}$, $D_r^{\omega\,{\bf G'}}$ are complex
adjustable parameters, and $C^{i\omega_P\,{\bf G'}}=0$ for
$\omega=i\omega_P$.

\section{\label{sec:res} Results}

We determine the parameters $A^\omega$, $B_{rs}^\omega$, $C^{\omega\,{\bf
G'}}$, and $D_r^{\omega\,{\bf G'}}$ in Eqs.\ \eqref{eq:fit0} and
\eqref{eq:fit0wings} by a standard linear regression on values
$\chi_{0\,{\bf 0},{\bf 0}}({\bf q},\omega)$ and
$\chi_{0\,{\bf 0},{\bf G'}}({\bf q},\omega)$ computed for nonzero ${\bf
q}$-points inside a sphere of radius $q_{\rm c}$ centered in $\Gamma$.
We implement this procedure within the \verb;abinit; \cite{ABINIT,Gonze05}
package.
To test the effectiveness of the method, we apply it to the HEG in a simple
cubic cell geometry, and to bulk sodium and aluminum in their experimental
crystal structures (bcc $a=8.107\,a_0$, and fcc $a=7.652\,a_0$,
respectively).

\begin{figure}
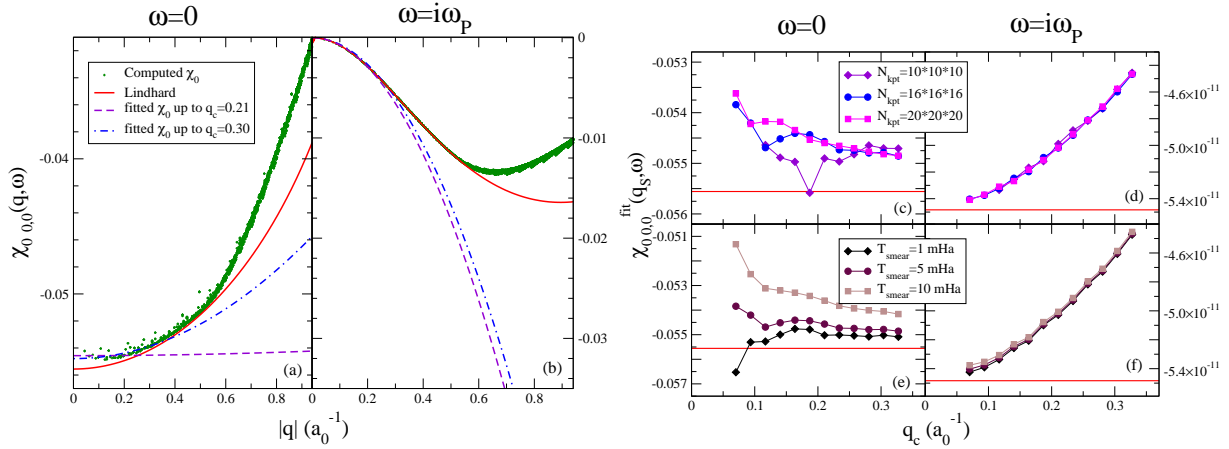

\includegraphics[width=8cm,clip=]{fit.eps}
\includegraphics[width=8cm,clip=]{conv.eps}
\caption{\label{fig:conv} (Color online)
The parabolic polarizability $\chi_0^{\rm fit}$, Eq.~\eqref{eq:fit0},
fitted to the computed $\chi_{0\,{\bf 0},{\bf 0}}({\bf q},\omega)$ of the
HEG (dots) restricted to ${\bf q}$-points within in a sphere of radius
$q_{\rm c}$ centered at ${\bf q}={\bf 0}$, and compared to the computed
polarizability itself and to the exact (Lindhard) function, for (a)
$\omega=0$ and (b) $\omega=i\omega_P$.
The computation involves a cut-off energy of 3~Ha,
$N_{\rm kpt}=16\times16\times16$ and a smearing temperature $T_{\rm
smear}=0.005$~Ha.
Panels (c)-(f): convergence of the fitted values $\chi_0^{\rm fit}({\bf
q}_s,\omega)$ (where the tiny ${\bf q}_s =(7,14,21) \,10^{-6}\, a_0^{-1}$)
as a function of the cutoff radius $q_{\rm c}$, for different ${\bf
k}$-points sampling, and with (c) $\omega=0$ and (d) $\omega=i\omega_P$,
and for different smearing temperature, and with (e) $\omega=0$ and (f)
$\omega=i\omega_P$.
Horizontal lines: the exact (Lindhard) values.
}
\end{figure}

Figure \ref{fig:conv} displays the fitting of the computed polarizability
$\chi_0$ of the HEG.
Panels (a) and (b) compare the computed $\chi_0$ and its small-${\bf q}$
fitted parabolic expansion, for two different cut-off radii $q_{\rm c}$.
Panels (c)-(f) display the resulting extrapolated small-${\bf q}$ values of
the polarizability as functions of the main parameters involved in the
simulations and the fit.
In these fits, the cut-off radius $q_{\rm c}$ cannot be chosen too small,
or else the number of ${\bf q}$-points becomes insufficient to perform a
reliable fit, especially at $\omega=0$, where the computed $\chi_0$ is
affected by significant numerical noise.
Likewise, if $q_{\rm c}$ is increased so much that it becomes comparable
with the Fermi momentum $k_{\rm F}$, the outer points introduce a
systematic error due to the non-parabolic ${\bf q}$-dependency of $\chi_0$.
An intermediate reasonably selected $q_{\rm c}$ must then be adopted.  This
is slightly more important for $\omega=i\omega_P$, where the fit is
comparably more sensitive to the value of $q_{\rm c}$, as shown in
Fig~\ref{fig:conv}(d,f).
Comparison to the Lindhard function show that this procedure provides a
fairly accurate small-$\bf q$ $\chi_0$ value, within few percent.
Small smearing temperature is beneficial to a better accuracy in the
determination of the asymptotic small-$\bf q$ behavior, but increase the
numerical noise in the computed $\chi_0$.

\begin{figure}
\includegraphics[width=8cm,clip=]{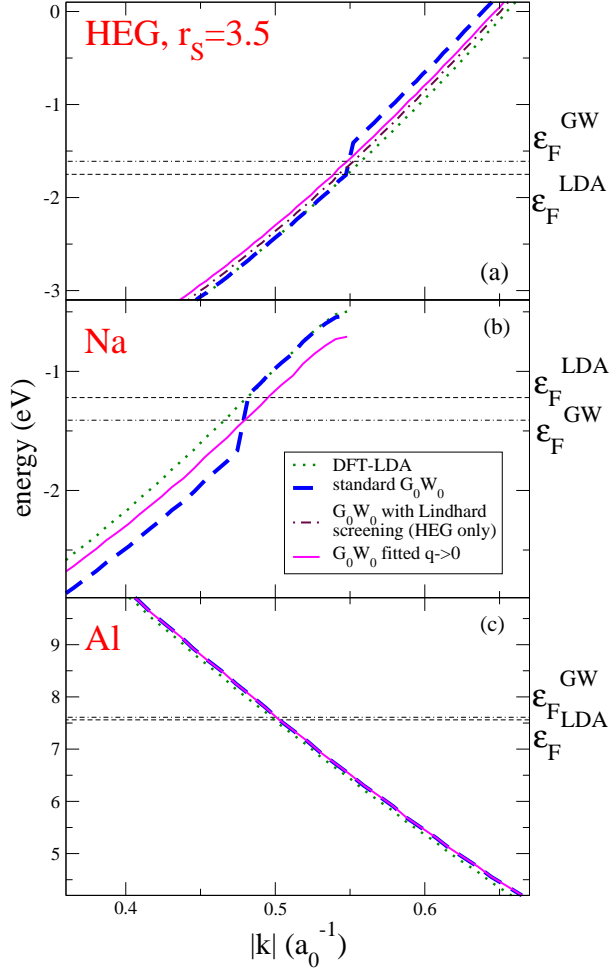}
\caption{\label{fig:fitted} (Color online)
Comparison of the band energy obtained via the naive $G_0W_0$ calculation
(dashed) to those obtained with the ${\bf q}\simeq 0$ corrected
polarizability (solid).
For the HEG (a) the figure also shows $G_0W_0$ results obtained with the 
analytic (Lindhard) polarizability.
For sodium (b) and aluminum (c) the bands are plotted along the ($110$) and
($11\overline{1}$) directions, respectively. }
\end{figure}

As Fig.\ \ref{fig:fitted} shows, the corrected screening successfully
closes the unphysical gap.
Of course in aluminum, where the fictitious gap is almost invisible, we see no
significant difference in the $G_0W_0$ corrections computed with and
without the fit.
The resulting curves are not very sensitive to the fit details, such as the
value  of $q_{\rm c}$, or $N_{\rm kpt}$.
For the HEG we can compare the obtained bands with those computed {\it via}
the Lindhard screening: the tiny almost uniform shift is due to the
truncation in the number of empty states included in the summations of Eq.\
(\ref{eq:chi}), which makes screening different in the large-${\bf q}$
region, as illustrated in Fig.\ \ref{fig:jel_eps}.

\begin{table}
\caption{\label{tab:bw}
Occupied bandwidth for the metals studied in this paper.
The present results are compared to similar calculations and experimental
data.
For the HEG, the DFT-LDA result coincides with the free-electron model
Fermi energy ${\cal E}_{\rm F}=E_{\rm Ha}\,(9\pi/4)^{2/3}(r_s/a_0)^{-2}$.
All energies are in eV.
}
\begin{ruledtabular}
\begin{tabular}{lccccc}
  &\multicolumn{3}{c}{HEG} &Na&Al\\
$r_s/a_0$&3&3.5&4&3.93&2.07\\
\hline
DFT-LDA&5.57&4.09&3.13&3.15&11.01\\
HEG $G_0W_0$ \footnotemark[1]&5.24&-&2.86&-&-\\
$G_0W_0$ for metals \footnotemark[2]&-&-&-&2.52&10.0\\
present work&5.14&3.84&2.86&2.81&10.03\\
experiment&-&-&-&2.65\footnotemark[3]&10.6\footnotemark[4]\\
\end{tabular}
\end{ruledtabular}
\footnotetext[1]{Calculations by Hedin \cite{Hedin}.}
\footnotetext[2]{Calculations by Northrup {\it et al.}\ \cite{nhl}.}
\footnotetext[3]{Experiment by Lyo and Plummer \cite{naexp}.}
\footnotetext[4]{Experiment by Levinson {\it et al.}\ \cite{alexp}.}
\end{table}

Table~\ref{tab:bw} reports the occupied bandwidths of the metals studied in
this work compared to previous calculations and experimental values.
The comparison with the DFT-LDA values shows the well-known bandwidth
reduction.  The results for the HEG are close to Hedin's computations
\cite{Hedin}, while in the case of Na and Al the numerical value are
comparable with data in the literature \cite{nhl} and approach the
experimental values.

\section{\label{sec:conc}Conclusions}

In this work we have shown that the standard $G_0W_0$ implementation of
calculation of quasiparticle-corrected bandstructures, a basic tool to
account for weak correlations in semiconductors and insulators, describes
metals correctly only in exceptional cases, like Al, where a substantial
part of the Fermi surface falls very close to a Brillouin zone boundary,
hence interband contributions make up for the missing intraband screening.
In general (like in the HEG and Na), the incorrect intraband contribution 
to the small-${\bf q}$ screening induces the opening of an unphysical gap
at the Fermi energy.

The proposed solution recovers the correct ${\bf q}\rightarrow 0$
polarizability by fitting a few small-${\bf q}$ computed values, and solves
this difficulty: the gap disappears, and the electron effective mass shows
the expected few percent increase.
This method requires a negligible computational cost, contrary to other
solutions based on Fermi-surface mapping.

An entirely different solution can be devised, which avoids the limitations
of the expansion (\ref{eq:exp}), and requires no fit altogether.
Since the small-${\bf q}$ polarizability needs to be computed at a ${\bf
q}_s$ much smaller than those generated by any practical ${\bf
k}$-sampling, it is possible to solve the KS equations on {\em two}
${\bf k}$-point meshes shifted from one another by ${\bf q}_s$, and then
apply directly Eq.\ (\ref{eq:chi}).
We tried this method for Na and for the HEG, and find that the typical
accuracies practically achievable in KS eigenvalues and wavefunctions allow
us to compute $\chi_0$ only for moderately large $|\bf q|$, of the order of
few percent of the $\bf k$-mesh spacing.
The use of such a not-so-small $\bf q_{\rm s}$ as a representative of the
$\bf q\to\bf 0$ limit would induce systematic errors in the calculation of
the $G_0W_0$ corrections.
The fit method is therefore practically preferable.

\begin{acknowledgments}
The authors acknowledge S. Caravati for providing the routines for
simulating the HEG within \verb;abinit;.
They also thank G.\ P.\ Brivio, R.\ Del Sole,
P.\ Garcia-Gonzales, M.\ Gatti, R.\ W.\ Godby,
V.\ Olevano, L.\ Reining, F.\ Sottile, and M.\ Verstraete
for useful discussion.
This work was funded in part by the EU's 6th Framework Programme through
the NANOQUANTA Network of Excellence NMP4-CT-2004-500198.
\end{acknowledgments}


\begin{thebibliography}{31}
\bibitem[{heg()}]{heg}
\bibinfo{note}{Interband contributions are identically zero in the HEG}.
\end{thebibliography}

\end{document}